\documentclass[sigconf]{acmart}
\usepackage{xcolor}
\usepackage{adjustbox}
\usepackage{multicol}
\usepackage{amsmath}
\usepackage{multirow}
\usepackage{graphicx}
\usepackage{pbox}
\usepackage{CJKutf8}
\usepackage{tabularx}

\newcommand\red[1]{\textcolor{red}{#1}}

\AtBeginDocument{%
  \providecommand\BibTeX{{%
    \normalfont B\kern-0.5em{\scshape i\kern-0.25em b}\kern-0.8em\TeX}}}

\acmConference[MM'24]{Proceedings of the 32th ACM International Conference on Multimedia}{October 28 - November 1,
  2024}{Melbourne, Australia.}
\acmBooktitle{Proceedings of the 32th ACM International Conference on Multimedia (MM '24), October 28 - November 1, 2024, Melbourne, Australia}
\acmPrice{15.00}
\acmISBN{978-1-4503-XXXX-X/18/06}

\usepackage{enumitem}
\usepackage{multirow}
\usepackage{array}
\usepackage{bm}

\newcolumntype{P}[1]{>{\centering\arraybackslash}p{#1}}

\usepackage[belowskip=2pt,aboveskip=0pt]{caption}

\copyrightyear{2024}
\acmYear{2024}
\setcopyright{acmlicensed}\acmConference[MM '24]{Proceedings of the 32nd ACM International Conference on Multimedia}{October 28-November 1, 2024}{Melbourne, VIC, Australia}
\acmBooktitle{Proceedings of the 32nd ACM International Conference on Multimedia (MM '24), October 28-November 1, 2024, Melbourne, VIC, Australia}
\acmDOI{10.1145/3664647.3681521}
\acmISBN{979-8-4007-0686-8/24/10}

\begin{document}

\title{MultiHateClip: A Multilingual Benchmark Dataset for Hateful Video Detection on YouTube and Bilibili}


\author{Han Wang}
\orcid{0009-0007-4486-0693}
\affiliation{
  \department{Information Systems Technology and Design}
  \institution{Singapore University of Technology and Design}
  \streetaddress{8 Somapah Road}
  \city{Singapore}
  \country{Singapore}
  \postcode{487372}
}
\email{han_wang@sutd.edu.sg}

\author{Tan Rui Yang}
\orcid{0009-0007-1325-5888}
\affiliation{
  \department{Information Systems Technology and Design}
  \institution{Singapore University of Technology and Design}
  \streetaddress{8 Somapah Road}
  \city{Singapore}
  \country{Singapore}
  \postcode{487372}
}
\email{ruiyang_tan@mymail.sutd.edu.sg}

\author{Usman Naseem}
\orcid{0000-0003-0191-7171}
\affiliation{
  \department{School of Computing}
  \institution{Macquarie University}
  \streetaddress{Balaclava Road}
  \city{Sydney}
  \country{Australia}
  \postcode{2109}
}
\email{usman.naseem@mq.edu.au}

\author{Roy Ka-Wei Lee}
\orcid{0000-0002-1986-7750}
\affiliation{
  \department{Information Systems Technology and Design}
  \institution{Singapore University of Technology and Design}
  \streetaddress{8 Somapah Road}
  \city{Singapore}
  \country{Singapore}
  \postcode{487372}
}
\email{roy_lee@sutd.edu.sg}
\authornote{Contact Author}

\renewcommand{\shortauthors}{Han Wang, Tan Rui Yang, Usman Naseem, \& Roy Ka-Wei Lee}



\begin{abstract}
Hate speech is a pressing issue in modern society, with significant effects both online and offline. Recent research in hate speech detection has primarily centered on text-based media, largely overlooking multimodal content such as videos. Existing studies on hateful video datasets have predominantly focused on English content within a Western context and have been limited to binary labels (hateful or non-hateful), lacking detailed contextual information. 
This study presents \textsf{MultiHateClip}\footnote{\url{https://github.com/Social-AI-Studio/MultiHateClip}} , an novel multilingual dataset created through hate lexicons and human annotation. It aims to enhance the detection of hateful videos on platforms such as YouTube and Bilibili, including content in both English and Chinese languages. Comprising 2,000 videos annotated for hatefulness, offensiveness, and normalcy, this dataset provides a cross-cultural perspective on gender-based hate speech. Through a detailed examination of human annotation results, we discuss the differences between Chinese and English hateful videos and underscore the importance of different modalities in hateful and offensive video analysis. 
Evaluations of state-of-the-art video classification models, such as \textit{VLM}, \textit{GPT-4V} and \textit{Qwen-VL}, on \textsf{MultiHateClip} highlight the existing challenges in accurately distinguishing between hateful and offensive content and the urgent need for models that are both multimodally and culturally nuanced. 
\textsf{MultiHateClip} represents a foundational advance in enhancing hateful video detection by underscoring the necessity of a multimodal and culturally sensitive approach in combating online hate speech.

\red{\textbf{Disclaimer: This paper contains sensitive content that may be disturbing to some readers.}}
\end{abstract}



\maketitle 



\section{Introduction}
The rapid expansion of social media has revolutionized the way information is shared, enhancing connectivity among users within both offline and online communities. However, these platforms have increasingly become conduits for the dissemination of hateful content that targets individuals or groups based on race, religion, gender, and other characteristics \cite{fortuna2018survey,schmidt2017survey}. The proliferation of online hate speech not only fosters discord among communities but also escalates to real-world violent hate crimes, underscoring the urgent need to identify and mitigate such content.

Current research on detecting hateful content has primarily concentrated on text-based analysis \cite{fortuna2018survey,schmidt2017survey}, with recent advancements extending to multimodal forms, such as memes \cite{sharma2022detecting, nayak2022multimodal, cao2023pro, hee2023decoding, cao2022prompting}. Yet, the field of hateful video detection remains underexplored, largely due to the lack of comprehensive datasets. Videos harness the synergistic potential of visual, auditory, and textual components to spread hate speech and offensive content, subsequently piquing the curiosity of select researchers~\cite{das2023hatemm,wu2020detection,alcantara2020offensive}. For instance. in a recent study, Das et al. ~\cite{das2023hatemm} constructed an English video dataset to facilitate hateful video classification. However, the existing studies have largely focused on English videos based on Western context, and these datasets could only facilitate simple coarse-grained binary classification (hateful or non-hateful) without diving into the fine-grained analysis of the hate video context, e.g., identifying target victims in the hate video.

To fill these gaps, our study introduces \textsf{MultiHateClip}, a multilingual short clip video dataset that facilitates a more nuanced and comprehensive exploration of multimodal hateful video content. This dataset compiled English short clips from YouTube, a global platform known for its vast user-generated content and diverse viewership, and Chinese short clips from Bilibili, a leading Chinese video-sharing website that caters to a younger demographic with a focus on animation, comics, and games (ACG) content. In this study, we specifically focus on gender-based hate video in the Western and Chinese cultural contexts, presenting an unprecedented cross-cultural perspective on hate speech in digital media. \textsf{MultiHateClip} contains 2,000 English and Chinese short videos, which not only enriches the understanding of hate speech's multifaceted nature but also marks the first initiative to construct a cross-cultural video dataset dedicated to examining hate speech, specifically targeting gender-related issues across Western and Chinese domains.

When constructing \textsf{MultiHateClip}, we utilize 80 hate-specific lexicons for each language to identify relevant videos on YouTube and Bilibili. Our annotation process involved a team of native-speaker annotators who were familiar with Western and Chinese popular cultures. Following \cite{davidson2017automated}; we differentiate hateful videos from offensive ones and task the annotators to categorize the videos into three distinct groups: \textit{hateful}, \textit{offensive}, or \textit{normal}. Videos identified as \textit{hateful} or \textit{offensive} were required to undergo additional annotation to identify segments containing such content, determine the target of hate speech (e.g., \textit{Woman}, \textit{Man}, \textit{LGBTQ+}, etc.), and specify the contributing modalities—whether \textit{visual}, \textit{auditory}, or \textit{textual}.

We conducted a thorough analysis on the \textsf{MultiHateClip} dataset and outlined key insights that could influence hateful video detection. We noted a surprisingly low frequency of hate videos on both YouTube and Bilibili, even after a comprehensive review of 10,000+ videos. Ultimately, we curated annotations for 1,000 videos per language, but only around 300+ of them were labeled as hateful or offensive. The analysis of the dataset revealed a consistent pattern: a disproportionate amount of gender-based hate video targeting women, reflecting broader issues gender discrimination in society. Further evaluation of modality contributions in these videos underscored the complexity of hate speech communication. For instance, 80.4\% of the hateful/offensive Chinese videos on Bilibili combined multiple modalities, such as visual, auditory, and textual modalities, to convey their hatefulness. The multimodal nature of hate emphasizes the need for an integrated approach that combines multiple modalities for a deeper understanding of hate speech dynamics.

To evaluate current models for hate video classification, we tested several state-of-the-art ones on \textsf{MultiHateClip}. In multiclass hate video classification, \textit{GPT-4V} excelled with a macro F1-score of 0.63 for English, while \textit{mBert $\odot$ MFCC $\odot$ Vivit} excelled with 0.50 for Chinese. These results underscore the importance of multimodal information, while highlighting critical limitations in current classification approaches: the challenge of differentiating between hateful and offensive content; the inadequacy of pre-trained models on non-Western cultural data; and the insufficient understanding of implicit hate.
These findings underscore the necessity for 
improved video classification models to tackle these identified weaknesses.

Our research contributions are summarized as follows:
\begin{enumerate}
    \item We constructe \textsf{MultiHateClip}, a multilingual dataset of hateful short clip videos. This dataset is enriched with detailed annotations for videos deemed hateful or offensive, detailing the specific segments with offensiveness, the targeted victims, and the modalities that contribute to the content's offensiveness. Such comprehensive annotations are designed to serve as a foundational resource for subsequent research.
    \item Our exhaustive examination of \textsf{MultiHateClip} unveils multilingual and cultural-specific characteristics and the critical role of multimodal inputs in hate video detection. These insights are crucial for improving hateful video detection methods, guiding the development of more effective models.
    \item We have critically evaluated existing video classification models, identifying key areas of weakness: the challenge in differentiating between hateful and offensive content, the limitations of pre-trained models concerning non-Western cultural data, and the insufficient understanding of implicit hate. These evaluations not only underscore existing gaps but also chart potential avenues for future research to enhance hateful video detection.
\end{enumerate}


\section{Related Work}

\textbf{Text-based Hate Speech Detection.} Extensive research has focused on detecting hate speech within textual content, producing a variety of datasets from platforms like Twitter \cite{zhang2018detecting}, Stormfront \cite{de-gibert-etal-2018-hate}, and Fox News \cite{gao2017detecting}. While many studies engage in binary classification (hate speech or not), works such as \cite{davidson2017automated} and \cite{founta2018large} attempt to distinguish between hateful, offensive, and normal speech. Other research targets specific forms of hate speech, such as misogyny~\cite{fersini2018overview,waseem-hovy-2016-hateful} and racism~\cite{waseem-2016-racist}. Warner and Hirschberg\cite{warner2012detecting} expand on this by categorizing hate speech into seven victim categories. However, these efforts primarily focus on text, leaving a gap in datasets and analysis for video content\cite{chhabra2023survey}.

\textbf{Video-based Hate Speech Detection.} In contrast to the abundant textual hate speech datasets, video-based datasets remain underdeveloped. For instance, \cite{alcantara2020offensive} and \cite{wu2020detection} introduce datasets comprising 400 Portuguese and 300 English YouTube videos, respectively. However, their limited sizes constrain the training of robust multimodal classification models. These studies primarily rely on textual analysis for classification. \cite{das2023hatemm} expands the field by compiling 1,083 BitChute videos into a dataset, although it focuses on English videos with binary hate classification, missing information like contributing modality. Refer to Table \ref{tab:previous_research} for a summary of relevant datasets, incorporating ours (\textsf{MultiHateClip}), in the realm of hateful video detection.

\textbf{Multimodal Model Fusion in Hate Speech Detection.} Traditional fusion techniques in hate speech detection often involve concatenating representation vectors from pre-trained unimodal models into a composite model, a method known as \textit{late fusion}. Recent studies have demonstrated efficacy in hateful video classification tasks~\cite{das2023hatemm}. Despite its effectiveness, emerging evidence highlights the potential of early fusion could also improve performance~\cite{9190246}. Vision-Language (VL) models, such as VideoBERT \cite{sun2019videobert}, ClipBERT \cite{lei2021less}, UniVL \cite{luo2020univl}, VLM \cite{xu2021vlm}, GPT-4V \cite{achiam2023gpt4} and Qwen-VL\cite{Qwen-VL}, demonstrate promise in video analysis tasks.


Addressing the gaps in video-based hate speech research, we introduce a dataset of 2,000 videos annotated for hatefulness, offensiveness, and normalcy, covering English and Chinese cultures. This dataset not only includes video labels but also detailed annotations on the hateful/offensive segments, target victim and contributing modalities. 
We will use late and early fusion models like VLM, GPT-4V, and Qwen-VL to evaluate classification performance.

\begin{table*}[t]
\centering
\small
\caption{Summary of datasets in hateful video detection. H:hateful, O:offensive, N:normal/non-hateful/non-offensive.}
\label{tab:previous_research}
\begin{tabular}{|c|c|c|c|c|c|c|}
\hline
\textbf{Work} & \textbf{Language} & \textbf{Size} & \textbf{Label} &\textbf{Hateful Segment} & \textbf{Targeted Victim} & \textbf{Contributing modality} \\ \hline \hline
OffVidPT \cite{alcantara2020offensive} & Portuguese & 400 & O, N & $\times$ & $\times$ & $\times$ \\ \hline
Hate\_speech\_ dataset\_videos \cite{wu2020detection} & English & 300 & H, N & $\times$ & $\checkmark$ & $\times$ \\ \hline
HateMM \cite{das2023hatemm} & English & 1083 & H, N & $\checkmark$ & $\checkmark$ & $\times$ \\ \hline
MultiHateClip & Chinese, English & 2000 & H, O, N & $\checkmark$ & $\checkmark$ & $\checkmark$ \\ \hline
\end{tabular}
\end{table*}

\section{MultiHateClip Dataset}
\subsection{Data Collection}

\textbf{Data Sources}: 
We collected videos from the social media platforms YouTube and Bilibili. Launched in 2005, YouTube has emerged as the leading global video-sharing platform, boasting 2.7 billion monthly active users \cite{youtube2024}, and is known for its comprehensive content moderation policies \cite{trujillo2020bitchute}. Bilibili, established in 2010, has become a central hub for the Chinese animation, comics, and gaming community, with over 3.36 billion monthly active users \cite{bilibili2024}, offering a unique insight into Chinese digital culture.


\begin{CJK*}{UTF8}{gbsn}
\textbf{Video Collection}: To source videos, we curated 80 pairs of gender-based hate lexicons based on \cite{bassignana2018hurtlex} and \cite{jiang2022swsr}, targeting synonymous terms across English and Chinese, such as 'mistress' and '情妇'. Using the YouTube and Bilibili APIs, we conducted keyword searches from Jan 24 to Feb 4, 2024, targeting short clips no longer than 60 seconds to focus on brief, potentially virulent content. This approach resulted in the collection of 5,600 English and 5,100 Chinese videos. Acknowledging the platforms' content moderation policies, which limit the presence of hate speech, we employed ChatGPT-3.5\footnote{\url{https://api.openai.com/v1/chat/completions}} to conduct an initial categorization based on video titles and transcripts. This step aimed to sift through the amassed videos, singling out those potentially featuring hateful or offensive content for closer examination. Ultimately, 2,000 videos from each language were selected for detailed manual annotation, ensuring a comprehensive analysis of hate speech trends within the dataset.
\end{CJK*}


\subsection{Human Annotation}

\textbf{Annotation Guidelines}. 
Our annotation process requires annotators to address four key questions for each short clip, aimed at comprehensively assessing and categorizing its content.

\textbf{Q1) Video Category Labeling}: Annotators are asked to classify each video into one of the three categories: \textit{Hateful}, \textit{Offensive}, \textit{Normal}. The category definitions are provided to guide the annotators:
\begin{itemize}
    \item \textit{Hateful}: Videos that incite discrimination or demean individuals or groups based on attributes such as race, ethnicity, nationality, religion, disability, age, veteran status, sexual orientation, gender identity, etc.
    \item \textit{Offensive}: Videos that may cause discomfort or distress, yet do not qualify as hateful under the criteria defined above.
    \item \textit{Normal}: Content devoid of hatefulness or offensiveness.
\end{itemize}


\textbf{Q2) Identification of Hateful/Offensive Segment}: For videos classified as hateful or offensive, annotators are tasked with determining the precise segment containing such content. They must specify the start and end times of the segment where the hateful or offensive statements occur. This requirement ensures a targeted analysis of the content, facilitating a more detailed examination of the nature and context of hate speech within the video.




\textbf{Q3) Identification of the Target Victim}: For videos identified as hateful or offensive, annotators are required to determine the target of the content. Given the dataset’s focus on gender-related hate speech, annotators should specify the intended victim group—\textit{Man}, \textit{Woman}, or \textit{LGBTQ+}. Additionally, there is an option to identify other groups if the video targets individuals or communities outside these categories. This step ensures a nuanced understanding of hate speech targeting, allowing for a comprehensive analysis of the videos' impact on various demographics.


\textbf{Q4) Determination of Contributing Modality}: In this part of the annotation process, annotators are asked to identify which modality—or modalities—of the video contribute to its hateful or offensive nature. They will classify the contribution as stemming from one or more of the following three categories: \textit{text} (including titles and transcripts), \textit{visual} content, or \textit{audio} content. This step is crucial for understanding how hate speech is conveyed through different channels within a video, whether through written words, visual imagery, or spoken language, offering insights into the multimodal dynamics of hate speech dissemination.



\textbf{Annotators Recruitment and Training}. 
Two PhD students, proficient in hate speech, served as expert annotators, supported by a team of 18 undergraduate students. All annotators are Asian individuals aged 18-24. The undergraduate team was balanced in terms of language proficiency—split evenly between the two languages of the study—and gender representation within each language group.

Prior to the annotation, annotators underwent training to grasp the guidelines and procedures. They annotated a test set of 30 videos across three rounds to assess their adherence to the guidelines. Expert annotators reviewed these annotations to ensure accuracy and provide guidance on errors.



\textbf{Annotators Process}. The annotation process was designed for robustness: each video initially received annotations from two different annotators. In instances of disagreement regarding the categorization (hateful, offensive, or normal), a third annotator was enlisted to provide an additional perspective. If disagreements persist, the matter is escalated to the expert annotators for final annotation. The ultimate categorization of each video was determined through a majority vote, ensuring a high level of consensus and reliability in the annotated dataset.

To manage the workload efficiently, we implemented a batch annotation strategy, allocating approximately 30 videos to each annotator daily. This task was designed to be manageable within a 30 to 40-minute commitment per day. Expert annotators played a crucial role in quality control, examining the day's annotations to verify label distribution and conducting random checks on selected video annotations. This daily review process allowed for immediate feedback and rectification of any misconceptions, with particular attention paid to outliers or unexpected labeling trends. Annotators periodically discussed ambiguities to ensure a consistent understanding and application of the guidelines.



\begin{table}[t]
\centering
\small
\caption{Statistics of English YouTube videos. H:hateful, O:offensive, N:normal, T:total.}
\begin{tabular}{|l|c|c|c|c|}
\hline
 & \textbf{H} & \textbf{O} & \textbf{N} & \textbf{T}  \\ \hline \hline
Count& 82& 256& 662& 1000\\ \hline
Avg. Title len& 8.17& 8.40& 8.78& 8.63\\ \hline
Avg. Transcript len& 85.24& 67.25& 75.87& 74.43\\ \hline
Avg. Video len(sec)& 37.14& 31.64& 34.20& 33.78\\ \hline
Avg. H/O segment(sec) & 33.06& 23.84& -& -  \\ \hline
\end{tabular}
\label{tab:youtube_stats}
\end{table}

\begin{table}[t]
\centering
\small

\caption{Statistics of Chinese Bilibili videos. H:hateful, O:offensive, N:normal, T:total.}
\begin{tabular}{|l|c|c|c|c|c|}
\hline
 & \textbf{H} & \textbf{O} & \textbf{N} & \textbf{T}  \\ \hline \hline
Count& 128& 194& 678& 1000\\ \hline 
Avg. Title len& 7.52& 7.85& 8.79& 8.44\\ \hline
Avg. Transcript len& 20.27& 21.30& 45.95& 37.88\\ \hline
Avg. Video len(sec)& 26.85& 30.19& 33.17& 31.78\\ \hline
Avg. H/O segment(sec) & 24.82& 27.20& -& -  \\ \hline
\end{tabular}
\label{tab:bilibili_stats}
\end{table}


\begin{table}[t]
\centering
\small
\caption{Distributions of videos targeting each victim group. H:hateful, O:offensive.}
\begin{tabular}{|c|c|c|c|c|}
\hline
& \multicolumn{2}{|c|}{English} & \multicolumn{2}{|c|}{Chinese}  \\ \hline
\textbf{Victim} & \textbf{H} & \textbf{O} &  \textbf{H} & \textbf{O} \\ \hline \hline
Woman & 40 & 121 & 59 & 97 \\ \hline
Man & 28 & 60& 54  & 50 \\ \hline
LGBTQ & 35 & 30 & 32  & 29 \\ \hline
Others & 17  & 34 & 48 & 13 \\ \hline 
\end{tabular}
\label{tab:target_victim}
\end{table}

\subsection{Data Statistics and Analysis}

We assessed annotation reliability using Cohen's kappa, yielding scores of 0.62 for English and 0.51 for Chinese in multiclass classification. Simplifying to a binary system (Hateful and Offensive combined), scores increased to 0.72 for English and 0.66 for Chinese. These Cohen's kappa values affirm moderate agreement among our annotators and underscore the robustness of our dataset.

The notable variance in kappa scores between Chinese and English stems primarily from annotators' extensive exposure to English social media. In English, 54 out of 227 inconsistent annotations confused hateful with offensive; in Chinese, 101 out of 254 showed similar confusion. Notably, due to the limited availability of Hateful/Offensive data, 54 and 101 instances underscored the challenge of distinguishing between Hateful and Offensive.

\textbf{Dataset Statistics}: Each video's transcript was obtained via Google Cloud Speech-to-Text\footnote{\url{https://cloud.google.com/speech-to-text}}. We then documented key metrics such as the number of videos, word counts in titles and transcripts, video lengths, and lengths of segments identified as hateful or offensive. These statistics are presented in Table~\ref{tab:youtube_stats} for English videos and Table~\ref{tab:bilibili_stats} for Chinese videos.

\textbf{Label Distribution}: \textsf{MultiHateClip} has a 3:7 ratio of \textit{hateful}/ \textit{offensive} to \textit{normal} videos, reflecting the platforms' strict content moderation policies. 
Interestingly, Normal videos generally have longer titles and transcripts, except for English Hateful videos, which have longer transcripts. Additionally, Hateful videos have a higher proportion of hateful segments compared to Offensive ones.

\textbf{Victim Group Analysis}: The distribution of hateful and offensive content targeting specific victim groups is detailed in Table~\ref{tab:target_victim}. Women were the most frequently targeted group across both languages and categories, followed by men and LGBTQ individuals, except in English Hateful videos, where men were more frequently targeted than LGBTQ individuals. A distinction emerges in the targeting of ``\textit{Other}'' victims: English videos more frequently target individuals based on religion or race, while Chinese videos tend to focus on nationality.

\begin{table}[t]
\centering
\small

\caption{Modality distribution of Hateful and Offensive videos. H:hateful, O:offensive.}
\begin{tabular}{|c|c|c|c|c|}
\hline
& \multicolumn{2}{|c|}{English} & \multicolumn{2}{|c|}{Chinese}  \\ \hline
\textbf{Modality} & \textbf{H} & \textbf{O} &  \textbf{H} & \textbf{O} \\ \hline \hline
Text & 12 & 97 & 19 & 38 \\ \hline
Audio & 1 & 2 & 0 & 0 \\ \hline
Vision & 0 & 12 & 0 & 6 \\ \hline
Text $\odot$ Vision & 7 & 46 & 40 & 63 \\ \hline
Audio $\odot$ Vision & 0 & 2 & 0 & 1 \\ \hline
Text $\odot$ Audio & 25 & 42 & 16 & 25 \\ \hline
Text $\odot$ Audio $\odot$ Vision & 37 & 55 & 53 & 61 \\ \hline
\end{tabular}
\label{tab:modality_contribution}
\end{table}

\subsection{Modality Analysis}
The contribution of different modalities to the hatefulness or offensiveness of content is summarized in Table~\ref{tab:modality_contribution}. Key findings include a higher prevalence of text-only hate content in English offensive videos and a significant portion of videos in both languages being classified as hateful or offensive due to multiple modalities. This underscores the importance of a multimodal approach in effectively identifying and classifying hate speech content.


\begin{CJK*}{UTF8}{gbsn}
\begin{table}[t]
\centering
\small
\caption{Top 10 highest tf-idf score keywords in videos' transcripts and titles. with Hateful keywords highlight in \red{red}. H:hateful, O:offensive, N:normal}
\begin{tabular}{|c|c|c|c|c|c|}
\hline
\multicolumn{3}{|c|}{English} & \multicolumn{3}{|c|}{Chinese}  \\ \hline
\textbf{H} & \textbf{O} & \textbf{N} & \textbf{H} & \textbf{O} & \textbf{N} \\ \hline 
\hline

like& like& like& \textcolor{red}{娘炮}& \textcolor{red}{娘炮}& 一个\\ \hline

that& \textcolor{red}{whore}& gay& 一个& 公主& \textcolor{red}{阴茎}\\ \hline
shorts& know& shorts& 英文& 一个& 花痴\\ \hline
oh& man& know& 仙女& \textcolor{red}{鸡脚}& 男人\\ \hline
going& that& that& 19& 露出& 时间\\ \hline
\textcolor{red}{faggy}& shorts& oh& covid& \textcolor{red}{泼妇}& 方法\\ \hline
yeah& oh& yeah& 日本& \textcolor{red}{婊子}& 分钟\\ \hline
know& going& \textcolor{red}{harpy}& 有没有& \textcolor{red}{贱人}& 妈妈\\ \hline
okay& \textcolor{red}{pussy}& get& 公主& 男人& \textcolor{red}{他妈的}\\ \hline
women& \textcolor{red}{faggy}& going& 真的& \textcolor{red}{老鸨} & 少妇
\\ \hline
\end{tabular}
\label{tab:tf_idf_results}
\end{table}
\end{CJK*}

\begin{figure*}[t] 
	\centering
	\includegraphics[scale = 0.35]{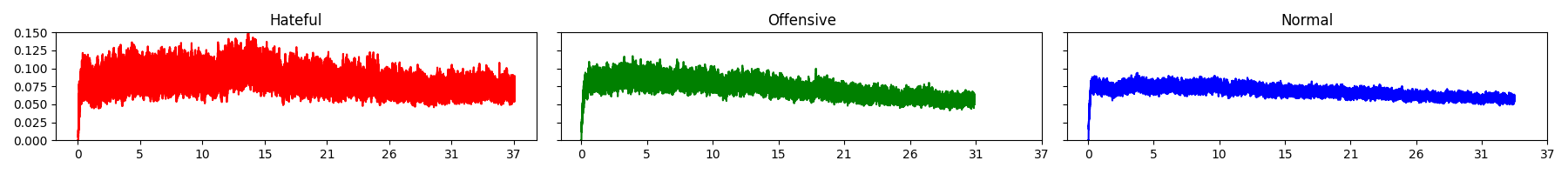} 
	\caption{Amplitude of English YouTube videos. Y-axis: Amplitude Indicator, X-axis: Time(sec.)}
	\label{fig:en_amplitude}
\end{figure*}

\begin{figure*}[t] 
	\centering
	\includegraphics[scale = 0.35]{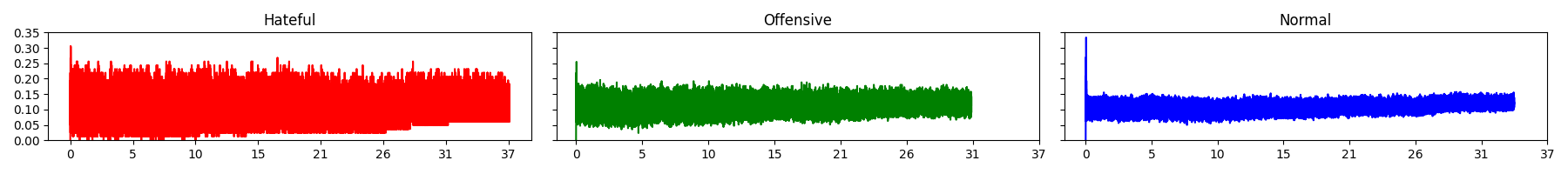} 
	\caption{Zero Crossing Rate of English YouTube videos. Y-axis: Zero Crossing Indicator, X-axis: Time(sec.)}
	\label{fig:en_zero_crossing_rate}
\end{figure*}

\textbf{Text Analysis}: 
We first combine video titles and transcripts into a unified text feature, with stop words removed for clarity. For Chinese text, the Jieba Python library\footnote{\url{https://pypi.org/project/jieba/}}  was employed for sentence segmentation. A term frequency-inverse document frequency (tf-idf) analysis yielded insights into the most prevalent words across categories, detailed in Table~\ref{tab:tf_idf_results}. 
Offensive videos in both languages predominantly feature hate lexicons. Nonetheless, distinguishing between Hateful and Normal videos based solely on text presents challenges, indicating the potential inadequacy of text alone for hate video detection.

\textbf{Audio Analysis}: Our evaluation of audio content focused on amplitude and Zero Crossing Rate (ZCR)  metrics. Amplitude analysis of English videos, represented in Figure~\ref{fig:en_amplitude}, highlighted that \textit{Hateful} and \textit{Offensive} typically exhibit higher sound intensities. Similarly, ZCR analysis of English videos, shown in Figure~\ref{fig:en_zero_crossing_rate}, indicated that these videos also feature higher rates of zero crossing, pointing to a potentially noisier audio profile. These consistent patterns across languages highlight the audio modality's role in video classification.

\textbf{Vision Analysis}: In our visual data analysis, we extracted one frame per second from each video, applying YOLOv3 for object detection on these frames \cite{redmon2018yolov3}. This analysis revealed a notable trend: English videos categorized as Hateful and Offensive had a higher detection rate of 'person' objects (70\% and 68\%, respectively) compared to Normal videos (63\%). Conversely, the application of YOLOv3 on Chinese videos resulted in a significant number of detection failures—704 instances for Chinese videos, compared to 289 for English. This discrepancy highlights a crucial limitation of current object detection models, which may stem from the underrepresentation of diverse gender and racial characteristics in training datasets like Coco\cite{lin2014microsoft,zhao2021understanding}. Such disparities pose substantial challenges in accurately recognizing individuals in videos from non-Western contexts, thereby impacting the reliability of our visual analysis results for Chinese content.

\section{Benchmarking Models}
In this section, we outline the approach for detecting hatefulness in videos using the \textsf{MultiHateClip} dataset and describe the benchmarking of various models tailored for this purpose.

\subsection{Problem Definition}
The task is to classify a given video \(V\) into one of three categories: Hateful (\(y=0\)), Offensive (\(y=1\)), or Normal (\(y=2\)). This classification considers three modalities within the video: \textbf{Text} (\(T\)): Comprising words \(\{w_1, w_2, \ldots, w_m\}\), aggregated from the video's title and transcript. \textbf{Audio} (\(A\)): The auditory component of the video. \textbf{Vision} (\(V\)): A sequence of frames \(\{f_1, f_2, \ldots, f_n\}\) extracted from the video.

The challenge for the model \(f\) is to accurately map these modalities to the ground-truth label \(y\), where \(f: f(T, A, V) \rightarrow y\) and \(y\) belongs to the set \(\{0, 1, 2\}\). While initially treated as a multiclass classification problem, we also consider a binary classification scenario by combining Hateful and Offensive categories into a single label for a simplified offensive label analysis.

\subsection{Preprocessing}
For text data, preprocessing involves removing stop words. For vision-based analysis, we select one frame per second and use padding images to standardize the analysis to 60 frames for videos shorter than 60 seconds.

\subsection{Text-Based Models}

\textbf{mBERT:} 
We use mBERT \cite{devlin2018bert} for its effectiveness in hate speech detection \cite{mozafari2020bert} and low-resource languages \cite{velankar2022mono}. Text data is processed with mBERT to extract 768-dimensional features, which are classified into Hateful, Offensive, or Normal via two FC layers. This setup is called Model T1.

\textbf{GPT-4-Vision-Preview without vision (GPT-4):} 
In our experiment with \textit{GPT-4-Vision-Preview} \cite{achiam2023gpt4}, we focus solely on linguistic analysis, excluding vision features. We prepare 4 demonstration examples with video text, a classification query, and the actual label, as shown in Table~\ref{tab:gpt_4v_classification}. We then use \textit{GPT-4} to predict labels for new video prompts, calling this setup Model T2.

\textbf{Qwen:}
Qwen-VL-7B \cite{Qwen-VL}, Alibaba Cloud's visual multimodal model, excels in vision-language tasks like captioning and VQA. We use a GPT-4-like prompt for label prediction, calling this setup Model T3.

\subsection{Audio-Based Models}
\textbf{MFCC:} 
Mel Frequency Cepstral Coefficients (MFCC) are crucial in audio signal processing and effective for audio classification \cite{xu2004hmm,vimal2021mfcc,banuroopa2021mfcc}. We generate a 128-dimensional MFCC vector for each audio signal and classify it into Hateful, Offensive, or Normal using two FC layers. This model is called A1.

\textbf{Wav2Vec2-BERT (Wav2Vec):}  
The Wav2Vec2-BERT model \cite{barrault2023seamless}, pre-trained on 4.5 million hours of audio in 143+ languages, excels in tasks like ASR and audio classification. We extract 1024-dimensional audio features with Wav2Vec2-BERT and use two FC layers for label prediction. This model is called A2.

\subsection{Vision-Based Models}

\textbf{ViViT:} 
The Video Vision Transformer (ViViT) \cite{arnab2021ViViT}, effective in anomaly \cite{yuan2021transanomaly} and violence detection \cite{singh2022video}, extracts a 768-dimensional feature vector from 32 sampled frames per video. These are classified through two FC layers, referred to as Model V1.

\textbf{ViT:} 
The Vision Transformer (ViT) \cite{dosovitskiy2020image}, excels in image recognition and detecting offensive content \cite{nayak2022multimodal}. In our study, ViT processes 60 video frames, generating 768-dimensional feature vectors. These vectors are classified using an LSTM network \cite{hochreiter1997long} followed by two FC layers, referred to as Model V2.

\subsection{Vision-Language Models}
\textbf{VLM:} 
The Vision-Language Model (VLM) \cite{xu2021vlm}, using task-agnostic multimodal pre-training, excels in video captioning and retrieval \cite{ruan2022survey}. In our study, VLM extracts two 768-dimensional representations from text and visual inputs, processes them through separate FC layers, then a fusion FC layer, and finally a classification FC layer for video detection. This method is called Model VL1.

\textbf{GPT-4-Vision-Preview (GPT-4V):} 
Specifications for the GPT-4V model are in Table~\ref{tab:gpt_4v_classification}. This model enhances \textit{GPT-4} by integrating visual inputs. In demonstration examples, visual inputs are summarized by GPT-4 with the prompt, “\textit{Please summarize the video's content}.” For new videos, we sample 4 * N frames and combine four frames into a single image to increase information density. Both demonstration examples and video prompts are processed by GPT-4V to predict classification labels, designated as Model VL2.

\textbf{Qwen-VL:}
The prompt format follows the GPT-4V model, but utilizes the Qwen-VL model. This model is designated as VL3.

\subsection{Multimodal Models}
    
\begin{figure}[t] 
	\centering
	\includegraphics[scale = 0.35]{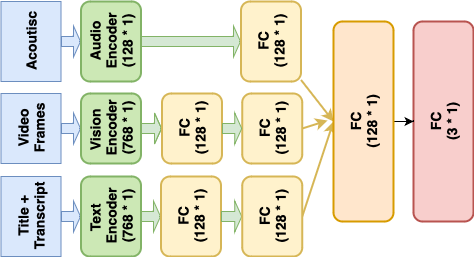} 
	\caption{Framework of the multi-modal model. FC: Fully Connected Layer.}
	\label{fig:framework1}
\end{figure}

\textbf{T1 $\odot$ A1 $\odot$ V1:} As shown in Figure~\ref{fig:framework1}, this model uses a fusion strategy combining mBERT for text, ViViT for visual features, and MFCC for audio, chosen for their superior performance within their modalities. Each set of features is processed through separate FC layers, then concatenated and classified through an additional FC layer. This integrated approach is called Model M1.

\definecolor{question_color}{HTML}{1B9E77}
\definecolor{label_color}{HTML}{D95F02}
\definecolor{explanation_provided_color}{HTML}{7570B3}
\definecolor{explanation_generated_color}{HTML}{FF55A3}
\definecolor{tweet_color}{HTML}{FFC300}

\begin{table}[htbp]
\centering
\small
\caption{Example GPT-4V's prompt for \textit{Hateful Video Classification} (reformatted for visualisation purposes). The prompt input consists of the extracted textual features (in \textcolor{explanation_provided_color}{purple}), extracted vision features(in \textcolor{explanation_generated_color}{pink}), the question (in \textcolor{question_color}{green}), and the GPT-4V's generated answers (in \textcolor{label_color}{orange}).}
\begin{tabular}{p{0.90\linewidth}}
\toprule
\# Demonstration Examples $1 ... 4$\\
\textbf{Input}: Video with Title: \textcolor{explanation_provided_color}{"POV: GTA Players when they see a hooker"}, Transcript: \textcolor{explanation_provided_color}{"Do they always do that? Shit? Oh. Do we have here?..."} and Vision Content: \textcolor{explanation_generated_color}{"The video appears to be a humorous and exaggerated depiction of a player's reaction..."}\\
\textcolor{question_color}{If the video is normal, output 'Normal'; if it is offensive, output 'Offensive'; if it is hateful, output 'Hateful'.} \\
\textbf{Output}: \textcolor{label_color}{Hateful}
\\
... 
\\
\# Actual Prompt\\
\textbf{Input}: Video with Title: \textcolor{explanation_provided_color}{"Mistress"}, Transcript: \textcolor{explanation_provided_color}{"What's the difference between wife and mistress about 50 pounds? later Wait, are you talking about a weight?"} and Vision Content as shown in images.\\
\makebox[3cm][l]{\textcolor{explanation_generated_color}{Image 1}} \makebox[3cm][l]{\textcolor{explanation_generated_color}{Image 2}}\\
\adjustbox{valign=m}{\includegraphics[width=3cm, height=5cm]{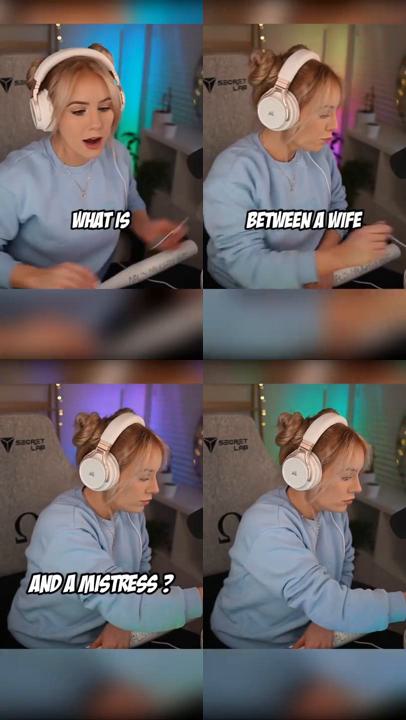}} 
\adjustbox{valign=m}{\includegraphics[width=3cm, height=5cm]{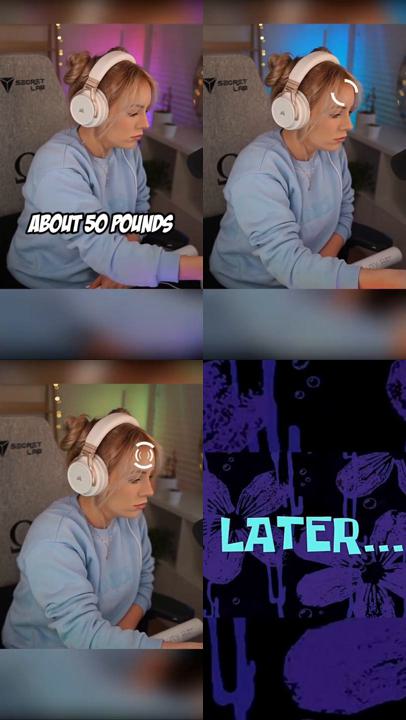}} \\
\makebox[3cm][l]{\textcolor{explanation_generated_color}{Image 3}} \makebox[3cm][l]{\textcolor{explanation_generated_color}{Image 4}}\\
\adjustbox{valign=m}{\includegraphics[width=3cm, height=5cm]{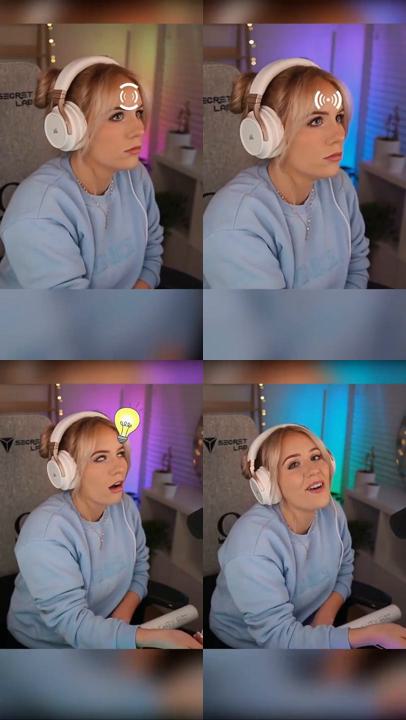}} 
\adjustbox{valign=m}{\includegraphics[width=3cm, height=5cm]{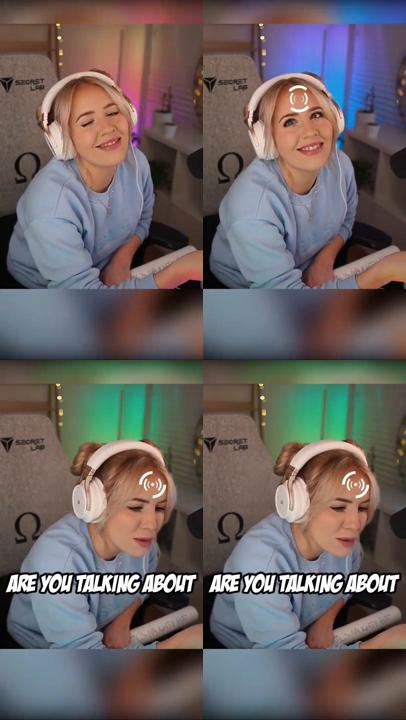}} \\
\textcolor{question_color}{If the video is normal, output 'Normal'; if it is offensive, output 'Offensive'; if it is hateful, output 'Hateful'.} \\
\textbf{Output}: 
\\
\bottomrule
\end{tabular}
\label{tab:gpt_4v_classification}
\end{table}


\section{Experiment}
\subsection{Experiments setup}

Our dataset was divided into training (70\%), validation (10\%), and testing (20\%) subsets, preserving label distributions. We detail each model's layer configurations below:


\begin{itemize}
    \item \textbf{mBERT:} 768-dimensional output passed through two FC layers (128 and 3 units).
    \item \textbf{MFCC:} 128-dimensional MFCC outputs input into two FC layers (128 and 3 units).
    \item \textbf{Wav2Vec:} 1024-dimensional MFCC outputs input into two FC layers (128 and 3 units).
    \item \textbf{ViViT:} 768-dimensional ViViT outputs input into two FC layers (128 and 3 units).
    \item \textbf{ViT:} Utilizes a 128-unit LSTM layer followed by two FC layers (128 and 3 units).
    \item \textbf{VLM:} Two 768-dimensional outputs undergo processing through two FC layers (128 units each), followed by fusion using one FC layer (128 units) and one FC layer for classification (3 units).
    \item \textbf{GPT-4V and Qwen-VL:} We experimented with prompt configurations using 2 to 16 images (N), finding N=4 optimal for both models in terms of accuracy and F1 scores.
\end{itemize}



For multimodal models, text and vision feature outputs are processed through two 128-unit FC layers, while audio feature outputs undergo processing through one 128-unit FC layer. These outputs are concatenated for final prediction through one 128-unit FC layer and one 3-unit FC layer. Each model's output, undergoes log-softmax transformation followed by cross-entropy loss calculation to estimate video label probabilities. Training spans 40 epochs using the Adam optimizer, with a batch size of 16 and a learning rate of 1e-4. Test performance is recorded when the model achieves its best Macro-F1 score on the validation set.

\subsection{Evaluation metrics}

To align with research norms, we use standard metrics: Accuracy, Macro F1 score, F1, precision, and recall, to evaluate our models on Hateful, Offensive, and Normal videos. For simplicity, Hateful and Offensive labels are combined into a single Offensive label for binary classification. The top metric is in \textbf{bold}, and the second-best is \underline{underscored}.

\subsection{Performance Evaluation}

Each model is run five times, except for GPT-4V and Qwen-VL related models, with results averaged. Model performance on multiclass classification tasks is outlined in Tables~\ref{tab:english_multiclass} (English videos) and~\ref{tab:chinese_multiclass} (Chinese videos).

The incorporation of multiple modalities notably boosts performance, as evidenced by comparisons such as M1 versus T1, A1, V1. To gauge the significance of these improvements, we calculate the p-value between the multimodal model (M1) and its best-performing unimodal counterpart (V1). Results show a significant enhancement for M1 compared to V1, attributed to the inclusion of text and audio modalities. For large vision language models like VL2 and VL3, integration of the vision modality generally enhances performance, with the exception of T3 outperforming VL3 in English data, possibly due to inconsistent performance of Qwen-VL, a Chinese model, on non-Chinese data. 
T2 and VL2 excel in English, highlighting the GPT-4V model's strength, while V1 and M1 lead in Chinese. Notably, VL2's top English performance surpasses M1's best Chinese performance by 13\% in Macro-F1 Score.

Further analysis of F1 scores for individual labels reveals some instances where the Hateful category obtains a score of zero, indicating challenges in distinguishing between Hateful and Offensive content, exacerbated by the low representation of Hateful videos in the dataset. This highlights the necessity of binary classification results for a fair comparison of model efficacy.

Binary classification results, detailed in Tables~\ref{tab:english_multiclass} for English and ~\ref{tab:chinese_multiclass} for Chinese videos. T2 and VL2 for English, and V1 and M1 for Chinese, consistently outperformed other models across most metrics. Furthermore, significant improvements were observed in evaluation metrics for the combined Offensive label across all three F1, Recall, and Precision metrics compared to initial separate Hateful and Offensive Label evaluations.

\begin{table*}[htbp]
  \centering
  \small
     \caption{Model performance for English YouTube hateful video classification. H:hateful, O:offensive, Acc:accuracy, M-F1:macro-F1, R:recall, P:precision, $^{*} p < 0.05$, $^{**} p < 0.01$, $^{***} p < 0.001$.}
  \begin{tabular}{ccccccccccccccc}
    \toprule
    & & \multicolumn{8}{c}{Multiclass}  & \multicolumn{5}{c}{Binary}\\
     &\textbf{Model} & \textbf{Acc} & \textbf{M-F1} & \textbf{F1(H)} & 
     \textbf{R(H)}  & \textbf{P(H)} & \textbf{F1(O)}  & \textbf{R(O)} & \textbf{P(O)}    & \textbf{Acc}  &\textbf{M-F1} & \textbf{F1(O)} & \textbf{R(O)}  & \textbf{P(O)}  \\
    \midrule
T1 & \textit{mBert} & 0.52& 0.35& 0.00& 0.00& 0.00& 0.43& 0.32& 0.67& 0.57& 0.57& 0.52& 0.42& 0.68\\  
 T2 & \textit{GPT-4} & \underline{0.76}& \underline{0.61}& \underline{0.35}& \underline{0.57}& \textbf{0.25}& \underline{0.64}& \underline{0.56}& 0.75& \textbf{0.81}& \textbf{0.79}& \textbf{0.73}& \underline{0.69}& 0.78\\  
   T3 & \textit{Qwen} & 0.64& 0.45& 0.1& 0.25& 0.06& 0.47& 0.38& 0.63& 0.72& 0.71& 0.65& 0.57& 0.75\\  
 A1 & \textit{MFCC} & 0.51& 0.32& 0.00& 0.00& 0.00& 0.33& 0.28& 0.41& 0.54& 0.50& 0.36& 0.33& 0.40\\  
 A2 & \textit{Wav2Vec} & 0.40& 0.27& 0.00& 0.00& 0.00& 0.51& 0.37& \textbf{0.83}& 0.53& 0.48& 0.64& 0.50& \textbf{0.90}\\  
 V1 & \textit{ViViT} & 0.66& 0.49& 0.12& 0.16& 0.10& 0.58& 0.45& \underline{0.80}& 0.73& 0.73& \underline{0.68}& 0.57& \underline{0.86}\\  
 
 V2 & \textit{Vit} & 0.59& 0.42& 0.18& 0.35& 0.14& 0.37& 0.37& 0.40& 0.63& 0.58& 0.44& 0.46& 0.45\\  
 
   VL1 & \textit{VLM} & 0.68& 0.42& 0.00& 0.00& 0.00& 0.47& 0.49& 0.45& 0.70& 0.64& 0.48& 0.59& 0.41\\ 
   
  VL2 & \textit{GPT-4V} & \textbf{0.77}& \textbf{0.63}& \textbf{0.36}& \textbf{0.67}& \textbf{0.25}& \textbf{0.66}& \textbf{0.60} & 0.73& \textbf{0.81}& \textbf{0.79}& \textbf{0.73}& \textbf{0.72}& 0.73\\

  VL3 & \textit{Qwen-VL} & 0.53& 0.35& 0.0& 0.0& 0.0& 0.4& 0.3& 0.61& 0.62& 0.61& 0.56& 0.46& 0.72\\ 
  
   M1 & \textit{T1 $\odot$ A1 $\odot$ V1} & 0.69$^{**}$& 0.54$^{**}$& 0.24$^{*}$& 0.27$^{*}$& 0.23$^{*}$& 0.59& 0.51$^{*}$& 0.71& \underline{0.75$^{*}$}& \underline{0.74}& 0.67& 0.61$^{*}$& 0.77\\ 
   
    \bottomrule
  \end{tabular}
 \label{tab:english_multiclass}
\end{table*}

\begin{table*}[htbp]
  \centering
  \small
  
    \caption{Model performance for Chinese Bilibili Hateful video classification. H:hateful, O:offensive, Acc:accuracy, M-F1:macro-F1, R:recall, P:precision, $^{*} p < 0.05$, $^{**} p < 0.01$, $^{***} p < 0.001$.}
  \begin{tabular}{ccccccccccccccc}
    \toprule
    & & \multicolumn{8}{c}{Multiclass}  & \multicolumn{5}{c}{Binary}\\
     &\textbf{Model} & \textbf{Acc}  & \textbf{M-F1} & \textbf{F1(H)} & 
     \textbf{R(H)}  & \textbf{P(H)} & \textbf{F1(O)}  & \textbf{R(O)} & \textbf{P(O)} & \textbf{Acc}    & \textbf{M-F1} & \textbf{F1(O)} & \textbf{R(O)}  & \textbf{P(O)}  \\
    \midrule

T1 & \textit{mBert} & 0.57& 0.46& \textbf{0.33}& 0.29& \underline{0.41}& 0.34& 0.29& 0.42& 0.67& 0.65& 0.58& 0.49& 0.71\\  
 T2 & \textit{GPT-4} & 0.58& 0.43& 0.14& \underline{0.50}& 0.08& 0.44& 0.30& \textbf{0.80}& 0.69& 0.68& \underline{0.63}& 0.51& \underline{0.84}\\  
  T3 & \textit{Qwen} & 0.51& 0.30& 0.00& 0.00& 0.00& 0.21& 0.26& 0.17& 0.56& 0.47& 0.26& 0.29& 0.25 \\ 
 A1 & \textit{MFCC} & 0.49& 0.38& 0.18& 0.15& 0.23& 0.30& 0.24& 0.38& 0.59& 0.58& 0.50& 0.41& 0.64\\  
 A2 & \textit{Wav2Vec} & 0.44& 0.36& 0.23& 0.15& \textbf{0.47}& 0.23& 0.24& 0.24& 0.57& 0.57& 0.52& 0.41& 0.73\\  
 V1 & \textit{ViViT} & 0.63& \underline{0.48}& 0.19& 0.16& 0.24& 0.44& 0.38 & 0.51& \underline{0.77}& \underline{0.76}& \textbf{0.71}& \underline{0.60}& \textbf{0.85}\\  
 V2 & \textit{Vit} & 0.56& 0.40& \underline{0.28}& 0.21& \textbf{0.47}& 0.17& 0.30& 0.13& 0.67& 0.64& 0.54& 0.50& 0.61\\
  VL1 & \textit{VLM} & 0.62 &	0.39 &	0.00 &	0.00 &	0.00 &	0.43 &	\textbf{0.46} &	0.45 &	0.64 &	0.58 &	0.42 &	0.46 &	0.38\\ 

 VL2 & \textit{GPT-4V} & \underline{0.66}& 0.47& 0.15& \textbf{1.00}& 0.08& \textbf{0.47}& 0.38& \underline{0.62}& 0.72& 0.69& 0.59& 0.59& 0.60\\  
VL3 & \textit{Qwen-VL} & 0.59& 0.40& 0.07& 0.33& 0.04& 0.39& 0.30& 0.55& 0.66& 0.62& 0.51& 0.47& 0.55 \\ 

 M1 & \textit{T1 $\odot$ A1 $\odot$ V1} & \textbf{0.68}& \textbf{0.50$^{*}$}& 0.20$^{**}$& 0.23$^{**}$& 0.20& \underline{0.46$^{**}$}& \underline{0.41$^{**}$}& 0.54$^{**}$& \textbf{0.80$^{**}$}& \textbf{0.78$^{**}$}& \textbf{0.71$^{**}$}& \textbf{0.66$^{*}$}& 0.78\\ 

    \bottomrule
  \end{tabular}
    \label{tab:chinese_multiclass}
\end{table*}

\definecolor{question_color}{HTML}{1B9E77}
\definecolor{label_color}{HTML}{D95F02}
\definecolor{explanation_provided_color}{HTML}{7570B3}
\definecolor{explanation_generated_color}{HTML}{FF55A3}
\definecolor{tweet_color}{HTML}{FFC300}


\begin{table*}[t]
  \centering
  \small
  
    \caption{Examples with Text, Visual Description, Ground Truth(GD), and Prediction. H:hateful, O:offensive, N:normal}
  \begin{tabular}{p{6.8cm}p{7cm}p{0.15cm}p{0.1cm}p{0.1cm}p{0.3cm}p{0.1cm}}
    \toprule
      \textbf{Text} & \textbf{Visual Description} & \textbf{GD} & \textbf{T1} & \textbf{V1} & \textbf{VL2}&  \textbf{M1}  \\ 
       \midrule
       Abuse caught on camera Husky kicked in front of a child. Houston SPCA and precinct 1 report owner kicking a dog...  & A woman kicked a dog, punched it and picked it up by one of its ears. All while a child was crying and begging her to stop.  & O & N & N & O & O\\ \hline
      Drag Queens Claim 'All White Are Racist' All white are racist. You can't be white and not racist. It's not possible because... & 5 Black drag queens converse animatedly. 1 queen stands, gestures angrily. Then, a white woman responds to the discussion.  & H & O & O & H & H \\ \hline
      
      Knock knock motherfuker. It's the USA. And I hear you guys got oil that isn't under my procession. So turn off the stove and return the oil to me. & First, an oil photo, followed by various images from a war movie depicting fighter planes, carrier trucks, and military command personnel. & O & O & O & O & H \\ \hline
      How 2 B. Kept Woman in the Dangerzone. Motivation for men to marry is low if wife is not going to have kids.... & A black man, attired in a black suit, exudes professionalism while delivering a speech at a microphone.  &H & N & N & H  & N  \\

    \bottomrule
  \end{tabular}
    \label{tab:error_analysis}
\end{table*}

\subsection{Error Analysis}

Table~\ref{tab:error_analysis} shows four English examples highlighting the importance of multimodal information and nuanced understanding. T1 and V1 make incorrect predictions in the first two examples, but VL2 and M1, using more modalities, predict correctly. Additionally, in 14 of 200 cases, T1 and V1’s unimodal predictions are wrong, whereas VL2 and M1’s multimodal predictions are accurate.

In the third and fourth examples, accurate prediction needs a deep understanding of video content. GPT-4V succeeds in both cases, while M1 fails. 
For the third example, M1's incorrect prediction likely stems from accumulating hatefulness across modalities, an issue also seen in the second example. Accurate hatefulness predictions require genuine content understanding, which M1 lacks. This issue resulted in 54 cases where M1 overpredicted hatefulness compared to T1, leading to 15 mispredictions.
In the fourth example, implicit hatefulness in text and vision components caused T1, V1, and M1 to misclassify the video as Normal. However, GPT-4V correctly identified it as Hateful. There are 11 similar cases where T1, V1, and M1 failed, but GPT-4V succeeded. These examples highlight GPT-4V's superior performance in understanding complex multimodal content.

\subsection{Limitations of Existing Models}
Our evaluation of the baseline models uncovers three significant challenges impacting their efficacy in hateful video detection:

\textbf{Distinguishing Between Hateful and Offensive Content:} 

Models often struggle to differentiate between Hateful and Offensive content, leading to frequent misclassifications. In the English data, 5 out of 11 models misclassify all Hateful videos. All models perform better at distinguishing Hateful/Offensive from Normal content. For example, the GPT-4V model has F1 scores of 0.36 for Hateful and 0.66 for Offensive, which increases to 0.73 when these labels are combined into a single Offensive label. This issue is also seen in the Chinese data, where models can separate Hateful/Offensive from Normal content but struggle to distinguish between Hateful and Offensive.
    
\textbf{Training Deficiencies on Non-Western Cultural Data:} 

Large-scale vision models like VL1, VL2, and the multimodal model M1 perform worse on Chinese datasets than on English ones due to limited non-Western cultural training data. As highlighted by \cite{vidgen2020directions}, there is a notable scarcity of non-Western hate speech datasets, with only 25 languages represented across 125 datasets, predominantly English (59 datasets). Chinese has only one documented dataset \cite{jiang2022swsr}, emphasizing the need for research in non-Western contexts. The recent emergence of Chinese hate speech datasets \cite{deng2022cold,lu2023facilitating} reflects growing interest in this area.
    

\textbf{Insufficient understanding of implicit content of current models.} GPT-4V excelled in English data primarily due to its extensive training data, enabling it to deeply understand both text and visual content. Certain videos, such as the fourth example in Table~\ref{tab:error_analysis}, contain implicit hate, necessitating large models capable of comprehending implicit content—an area where most hateful video classification models exhibit weakness.

These findings emphasize the necessity for tailored approaches in model training and modality fusion to effectively address the nuanced and culturally diverse nature of hate speech in videos.

\section{Conclusion}

This study introduces \textsf{MultiHateClip}, a multilingual dataset for hateful video detection, enriched with fine-grained labels across English and Chinese languages. Our investigation demonstrates the significant superiority of vision-language and multimodal models (\textit{GPT-4V} and \textit{mBert $\odot$ MFCC $\odot$ Vivit}) over unimodal models, highlighting the importance of modality integration. \textsf{MultiHateClip} not only provides a valuable resource for advancing this field but also underscores the critical necessity for multimodal analysis in understanding and combating hate speech in a multilingual context through extensive experimentation.

\section*{Acknowledgement}
This research/project is supported by Ministry of Education, Singapore, under its Academic Research Fund (AcRF) Tier 2. Any opinions, findings and conclusions or recommendations expressed in this material are those of the authors and do not reflect the views of the Ministry of Education, Singapore.


\bibliographystyle{ACM-Reference-Format}
\balance
\bibliography{ref}

\end{document}